\documentclass[aps,prd,preprint,groupedaddress,amssymb,amsmath,nofootinbib]{revtex4}
\usepackage[dvipsnames]{xcolor}
\usepackage{graphicx}
\usepackage{epstopdf}
\usepackage{hyperref} 

\begin{document}

\preprint{}

\title{Little hierarchy in the minimally specified MSSM}

\author{Radovan Derm\' \i\v sek }

\email[]{dermisek@indiana.edu}

\author{Navin McGinnis}

\email[]{nmmcginn@umail.iu.edu}

\affiliation{Physics Department, Indiana University, Bloomington, IN 47405, USA}



\date{May 4, 2017}

\begin{abstract}
We study constrained versions of the minimal supersymmetric model and  investigate the hierarchy between the electroweak scale and the scale of superpartners that can be achieved without relying on specifying model parameters by more than one digit  (or with better than 10\% precision). This approach automatically avoids scenarios in which a large hierarchy is obtained by special choices of parameters and yet keeps scenarios that would otherwise be disfavored by various sensitivity measures. We consider models with universal gaugino and scalar masses, models with non-universal Higgs masses or non-universal gaugino masses  
and focus on scenarios in which all the model parameters are either of the same order or zero  at the grand unification scale. We find that the maximal hierarchy between the electroweak scale and stop masses, requiring that model parameters are not specified beyond one digit, ranges from a factor of $\sim 10-30 $ for the CMSSM up to $\sim 300$ for models with non-universal Higgs or gaugino masses.

\end{abstract}


\maketitle






\section{Introduction}	

The hierarchy between the electroweak scale and the scale of new physics, with some scenarios  pushed well above the TeV scale, remains a mystery with respect to naturalness. Affected scenarios  include supersymmetric extensions of the standard model that otherwise have a number of attractive features leading to their popularity. While naturalness criteria do not rule out any model, nor are completely scientific, they shape our bias toward  theoretical models and even priorities for experimental searches and strategies.\footnote{There is a large number of papers written on the subject. Examples include Refs.~\cite{Barbieri:1987fn, Vissani:1997ys, Chan:1997bi, Barbieri:1998uv, Chankowski:1998xv, Giusti:1998gz, Romanino:1999ut, Baer:2012cf}. For reviews and further references see, for example, Refs.~\cite{Martin:1997ns, Giudice:2008bi, Dermisek:2009si, Hall:2011aa, Feng:2013pwa, Arvanitaki:2013yja}.}

Prevailing naturalness arguments are based on the largest contribution to an observable, in this case the electroweak scale,  and our intuition that two contributions should not cancel each other with a high precision unless there is a reason for it, like  a symmetry. Thus, if the  contribution of one parameter is much larger than the observed value, it is interpreted as the need for another parameter to be carefully adjusted in order to precisely cancel that contribution. Such a need for fine tuning of model parameters is considered unnatural.
For example, a typical contribution to the electroweak scale from scalar tops with masses of several TeV, required for the Higgs boson mass in the minimal supersymmetric model (MSSM),  is of order several TeV.  Since relevant quantities are masses squared, this contribution needs to be cancelled with precision of 1 part in $10^3 - 10^4$ in order to obtain the desired outcome. If there were just two model parameters contributing to the electroweak scale, this would indeed require fine tuning of the other parameter with 3 to 4 digit precision. However, it has been recently argued, that what is a very unnatural and fine-tuned outcome in a model with two parameters can be a completely ordinary outcome, not requiring carefully tuned parameters, in a model where more parameters are contributing to  given observable~\cite{Dermisek:2016zvl}.

In this paper, we adopt a top down approach and  investigate the hierarchy between the electroweak scale and the scale of superpartners (little hierarchy) that can be achieved without relying on specifying model parameters by more than one digit  (or with better than 10\% precision) in several constrained versions of the MSSM\footnote{We only discuss the hierarchy between the electroweak scale and the scale of superpartners, and not that of superpartners to higher scales, e.g. the GUT or Planck scales. The hierarchy between the scale of superpartners and a high scale is successfully addressed by supersymmetry.}.  We consider models with universal gaugino and scalar masses, models with non-universal Higgs masses or non-universal gaugino masses 
and focus on scenarios in which all the model parameters are either of the same order or zero  at the grand unification (GUT) scale. 
 This approach automatically avoids scenarios in which a large hierarchy is obtained by special choices of parameters  and yet keeps scenarios that would otherwise be disfavored by various sensitivity measures.
While  for simple models with 2 parameters the method leads to the same conclusions as other methods to estimate the scale of new physics that does not require fine tuning, for models with more parameters  the method leads to significantly weaker constraints on the scale of new physics and the possible little hierarchy grows with the complexity of the model.

In the following section we describe in detail the methodology we use on a simple  model with only the universal gaugino mass and the $\mu$-term. In Sec.~\ref{sec:results} we  find the maximal little hierarchy obtainable by requiring that model parameters are not specified beyond one digit in several constrained versions of the MSSM. We discuss the method further in Sec.~\ref{sec:discussion} and summarize results in Sec.~\ref{sec:conclusions}.

\section{Methodology}
\label{sec:method}

We focus on scenarios with soft supersymmetry breaking terms and  the supersymmetric Higgs mass, $\mu$, specified at the GUT scale. All the parameters of a given model  are, for simplicity, assumed to be either of the same order or zero at this scale. We label the common scale of all non-zero parameters by $M_{SUSY}$. In order to be specific, we allow parameters to vary by 50\% in both directions, in the interval $[0.5, 1.5]  M_{SUSY}$  for dimension mass parameters and $[0.5, 1.5]  M_{SUSY}^2$ for dimension mass-squared parameters,  and we specify the departures from the central value by one digit.  For example, $M_{1/2} = 0.8 M_{SUSY}$, $m_{0}^2 = 1.2 M_{SUSY}^2$, and so on where $M_{1/2}$ and $m_0$ are the universal gaugino and scalar partner masses at the GUT scale, respectively.  This one digit  procedure corresponds to selecting model parameters in 10\% intervals around the central values. Other choices could be made, e.g. 30\%, 1\% etc., and the obtained results could be ordered according to  the step in which model parameters are selected. We find the choice of 10\% interval as both reasonable and intuitive.

We  evolve all the parameters to the stop mass scale, defined as $m_{\tilde t}^2 = (m_{\tilde t_L}^2 + m_{\tilde t_R}^2)/2$, using two loop RG equations~\cite{Martin:1993zk} and obtain the electroweak scale, represented by the $Z$ boson mass, as a result of electroweak symmetry breaking (EWSB),
\begin{equation}
M_Z^2 \equiv -2|\mu|^2 - m_{H_u}^2 - m_{H_d}^2 + \frac{|m_{H_u}^2 - m_{H_d}^2|}{\sqrt{1-\sin^2 (2\beta)}},
\label{eq:MZ}
\end{equation}
where  $m_{H_{u,d}}^2$ are soft supersymmetry breaking masses squared of the two Higgs doublets and $\tan\beta$ is the ratio of vacuum expectation values of the two Higgs doublets.
 We plot the obtained hierarchy between $M_Z^2$ and $M_{SUSY}^2$ or $m_{\tilde t}^2$.  Since we are interested in the obtained hierarchy in a top-down approach starting with fixed $M_{SUSY}$,  the plotted $M_Z^2$  does not necessarily represent the correct mass of the $Z$ boson but rather  the right hand side of  Eq.~(\ref{eq:MZ}). We also plot the points when this quantity  is negative and the EWSB does not occur. Alternatively, but only for points where EWSB occurs, one could appropriately rescale $M_{SUSY}$ so that $M_Z$ is the correct mass of the $Z$ boson and get an indication for $M_{SUSY}$ required. We will follow  the former approach that is more suitable for a top-down approach and also gives a perception of the size of the parameter space where EWSB does not occur.

\begin{figure}[t]
\centering
\includegraphics[width=3.5in]{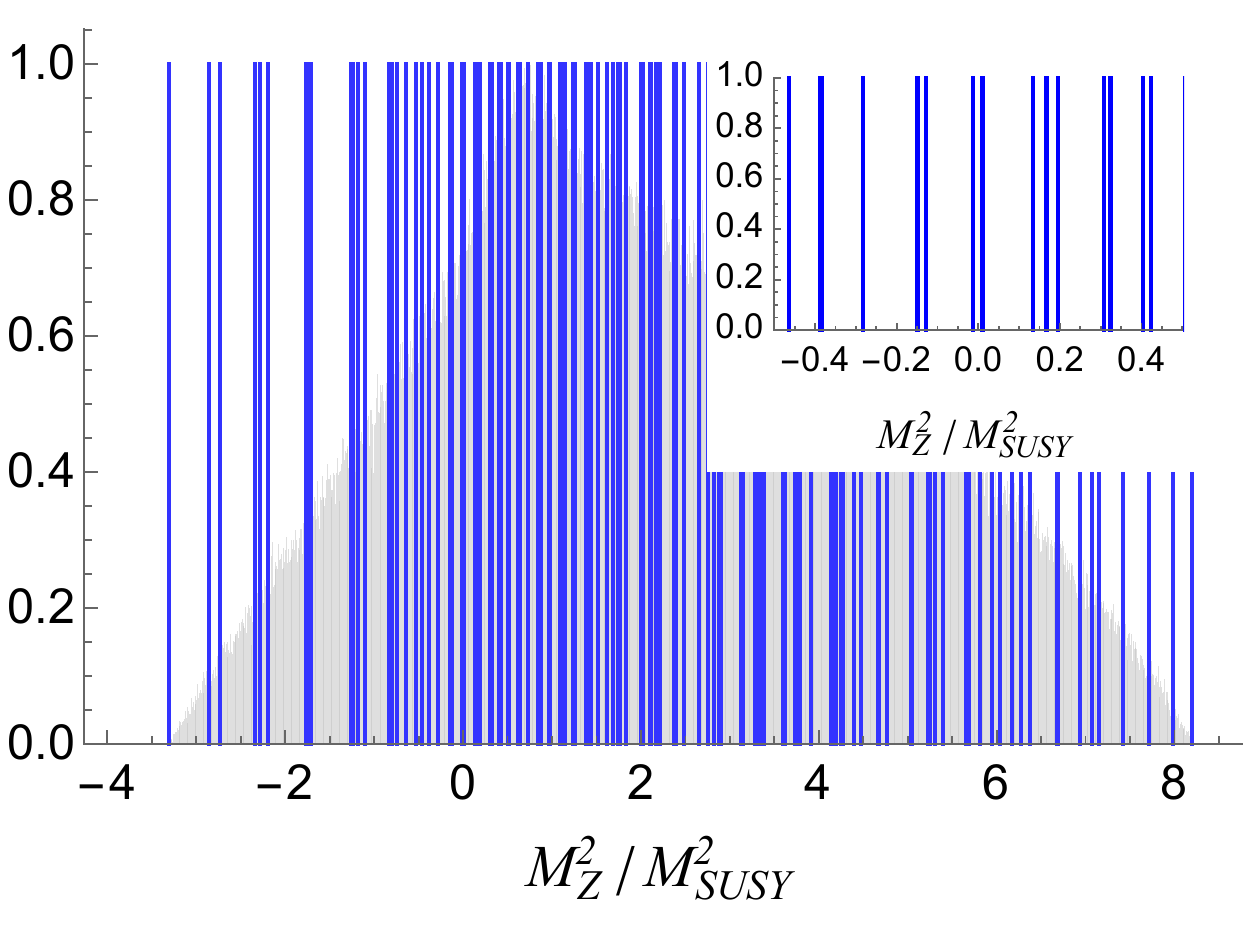}
\caption{The shaded area is the  distribution of $M_Z^2/M_{SUSY}^2$ for  randomly generated $\mu$ and $M_{1/2}$ in the interval $[0.5, 1.5]  M_{SUSY}$ for  $M_{SUSY} = 3$ TeV and $\tan\beta =10$   (arbitrarily normalized).  The highlighted lines (blue) are the outcomes of the grid scan described in the text with model parameters specified with one digit (or selected in 10\% steps from central values). The inset zooms in the region of small $M_Z^2/M_{SUSY}^2$. For negative values the EWSB does not occur, the values indicate the right hand side of Eq.~(\ref{eq:MZ}).} 
\label{fig:mu-M12}
\end{figure}

It is instructive to discuss further steps on a specific example. Let us consider a simple (although not  phenomenologically viable) model  with two non-zero mass parameters at the GUT scale, the $\mu$-term and the universal gaugino mass $M_{1/2}$.  If the $m_0$ was fixed  instead to a small value, sufficient to make the model phenomenologically viable, the results would be almost the same. The hierarchy obtained in our procedure is plotted with highlighted lines in Fig.~\ref{fig:mu-M12} for  $M_{SUSY} = 3$ TeV and $\tan\beta =10$ assuming $M_{GUT} = 3.2\times 10^{16}$ GeV.\footnote{Some choices of these parameters are needed for the numerical analysis. However, the results are essentially the same for any $\tan\beta > 5$ and vary slowly with $M_{SUSY}$. The specific value of the GUT scale does not have an appreciable effect of the results presented.}  The  lines  are the outcomes of the grid scan described above with model parameters specified with one digit  (or selected in 10\% steps from central values). For comparison, the shaded area corresponds to possible outcomes if both parameters were random real numbers in the same interval  (or selected in infinitely small steps around central values).

Now let us turn our attention to determining the smallest outcome that is guaranteed to occur in the distribution without being accidentally small as a result of special choices of central values or the intervals. The smallest outcome for $M_Z^2/M_{SUSY}^2$ that appears in Fig.~\ref{fig:mu-M12} is certainly not of interest. By slightly adjusting the central values of $\mu$, $M_{1/2}$, or both, one can shift the position of an outcome arbitrarily close to zero. Such an outcome would be accidentally small due to the careful choice of the model parameters. However, the gaps between the outcomes would not be sensitive to these adjustments since the parameters are still scanned in 10\% intervals around the central values. Regardless to how the central values are adjusted there will always be at least one outcome as small as the largest gap size. Therefore, it is the maximal gap size that indicates the smallest possible outcome guaranteed to appear in the distribution. This gives the largest possible hierarchy between $M_Z$ and $M_{SUSY}$that does not require any special choices of central values or the intervals.

 Neglecting outcomes smaller than the largest gap size also means that none of the model parameters needs to be specified with better than 10\% precision. As mentioned, repeating the same procedure for arbitrarily shifted (within 10\%) model parameters would lead to the same result for the largest gap. Since 10\% steps can be specified by 1 digit,  the independence of the result to an arbitrary shift within 10\% means that it does not matter what the remaining digits in all model parameters are. This intuitive connection is the main reason for choosing 10\% as the step size but, as mentioned before, any step size can be chosen and the obtained gap size is characteristic for a given step size. Clearly, a smaller step size (larger tuning) would lead to a smaller gap size (larger possible hierarchy between the EW scale and superpartners).

 Once we know what is the smallest outcome guaranteed to occur in the distribution  there is no logical argument that can be made to disfavor this outcome compared to any other specific outcome. Thus, it represents the smallest completely ordinary outcome that corresponds to selecting model parameters in given steps (or specifying them with given precision). We will continue this discussion in Sec.~\ref{sec:discussion}.

 Going back to our example, 
from the inset in  Fig.~\ref{fig:mu-M12}   we can see that the gaps between the outcomes are at most 0.13 from which we conclude that the smallest $M_Z$ among minimally specified  outcomes (outcomes resulting from specifying model parameters with one digit irrespectively of what the remaining digits are) is about 0.4$M_{SUSY} $. The much smaller outcome close to zero  in Fig.~\ref{fig:mu-M12} is accidental, it would move away if some of the inputs change. However, for large variations of input parameters, of order 1 in units of $M_{SUSY}$, some outcome is guaranteed to appear within the largest gap size from 0.  This is true no matter what the remaining digits of the input parameters are.

From a similar plot of $M_Z^2/m_{\tilde t}^2$ we would find the maximal gap size to be $\sim 0.11$ away from  edges of the distribution and thus we find $M_Z > 0.3 m_{\tilde t}$.\footnote{We will show such plots for realistic models in the following section. It is not important for illustrating the method.} Smaller outcomes, or larger hierarchy, can be obtained  if we specify the way parameters are varied with more digits. The gaps fill and arbitrarily large hierarchy can be achieved when fine tuning of parameters is allowed as indicated by the shaded area in Fig.~\ref{fig:mu-M12}.

These findings are in agreement with other methods to estimate the maximal hierarchy one obtains without tuning parameters.  Not allowing more than 10\% tuning with the usual naturalness measures, based on sensitivity to individual parameters,  requires superpartners not heavier than about  $300$ GeV in this model. The reason for the agreement is that, for a model with two parameters, less than  10\% tuning is essentially the same requirement as not specifying the input parameters with more than 1 digit. Two order one numbers do not cancel to a smaller number than 0.1 unless they are tuned to each other at more than 10\% level. Similarly, two order one numbers do not cancel to a smaller number than 0.1 regardless of what is the second digit specifying each parameter. However,  this equivalency  does not hold in models with more parameters as we will see in the following section.

 Our  method of estimating the range of outcomes for a given observable that does not require carefully adjusted model parameters is fairly independent of many assumptions and details of the analysis. As already mentioned, the results depend very little on the choice of $\tan \beta$  and vary slowly with $M_{SUSY}$.  Note however, that fixing $\tan \beta$ corresponds to choosing the soft supersymmetry breaking $b$-term, $\mathcal{L}_{\text{soft}} \supset - (bH_uH_d + c.c)$. We could  simply assume that  the $b$-term is in the range favoring EWSB and, since the results depend very little on $\tan\beta$, it does not have to be specified carefully. Furthermore, even if the $b$-term was treated as a free parameter and varied over the same region and specified with one digit as other parameters the results would be comparable. The difficulty with this approach is that there are points for which the right hand side of Eq.~(\ref{eq:MZ}) is not a real number and so these outcomes cannot be visualized easily. In addition, each point would have a different top Yukawa coupling which  is obscuring the meaning of the results somewhat (moreover, for points with no EWSB, there is no unique way to fix the top Yukawa coupling). Nevertheless, the procedure can be repeated for points where EWSB occurs and we found that the gap size shrinks by a factor of $\sim2$ compared to results with fixed $\tan \beta$. Thus, the results with fixed $\tan \beta$ can be viewed as conservative. As such, considering also the simplicity of the analysis and visualization of the results, we will follow the approach with fixed $\tan \beta$. 

Other details of the analysis also play a minor role. For example, stopping the RG evolution at a common scale, rather than $m_{\tilde t}$ scale specific for each point, does not make a significant difference.\footnote{Similarly, starting the RG evolution at a different scale than the GUT scale, for example the Planck scale or an intermediate scale has only a minor impact on the results. Note, however, that the results would be very different in models with superpartner masses generated at a very low scale. As the starting point of the RG evolution approaches $m_{\tilde{t}}$, the contributions of all parameters except $\mu$ and $m_{H_{u,d}}$ go to zero. The RG evolution for at least a few orders of magnitude in the energy scale allows for the contribution of additional parameters to the determination of the EW scale.} The results do not even depend on precise values of gauge and top yukawa couplings. Although we perform a precise analysis, the results would be very similar following a highly simplified procedure, even when specifying all couplings with one digit. The largest impact on the  gap size comes from the  interval over which the parameters are allowed to vary. Our choice of the interval, $[0.5, 1.5]  M_{SUSY}$, is not motivated by anything besides simplicity. It assumes that there is only one scale in the problem, $M_{SUSY}$, and the range is of the same order as the scale so that the results are robust to  ${\cal O}(1)$ changes in input parameters. It allows a hierarchy of 3 between the smallest and the largest parameter of the model. If parameters were allowed to vary over significantly  larger ranges the results would be affected. However, in such a case, it would make more sense to study separately two or more scale models with different hierarchies between parameters. We will not consider models with large hierarchies between non-zero parameters here.

\section{Results}
\label{sec:results}

In this section we present detailed results for the little hierarchy that is obtainable from minimally specified parameters in the constrained minimal supersymmetric model (CMSSM) characterized by the universal gaugino mass, $M_{1/2}$, universal scalar mass squared, $m_0^2$, and universal soft-trilinear coupling $A_0$ in addition to the supersymmetric Higgs mass, $\mu$, and $\tan\beta$. We also briefly explore models with more free parameters at the GUT scale, namely the models with non-universal Higgs masses, $m_{H_{u,d}}^2$, or non-universal masses of $SU(3)$, $SU(2)$, and $U(1)$ gauginos, $M_3$, $M_2$, and $M_1$, respectively.

\subsection{CMSSM}

\begin{figure}[t]
\centering
\includegraphics[width=2.6in]{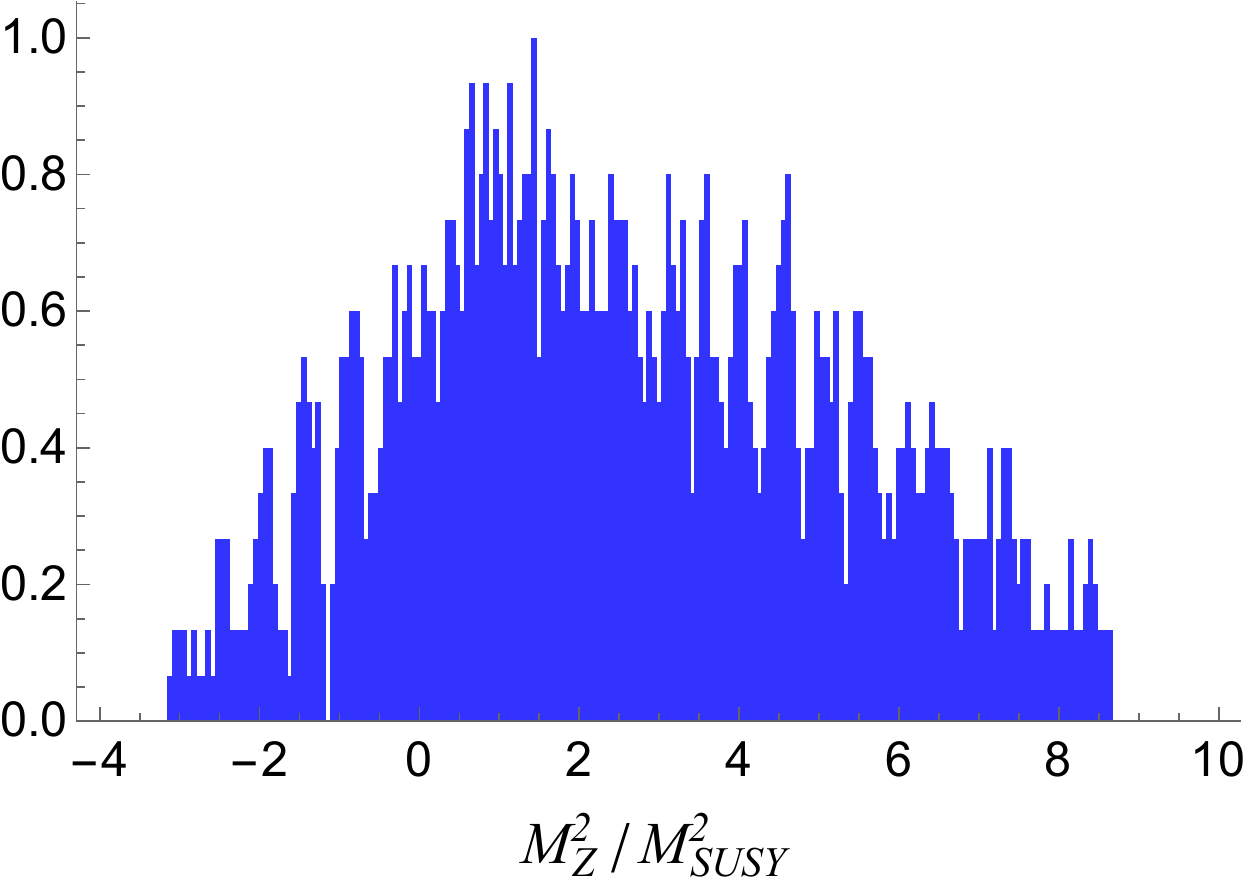}\includegraphics[width=2.6in]{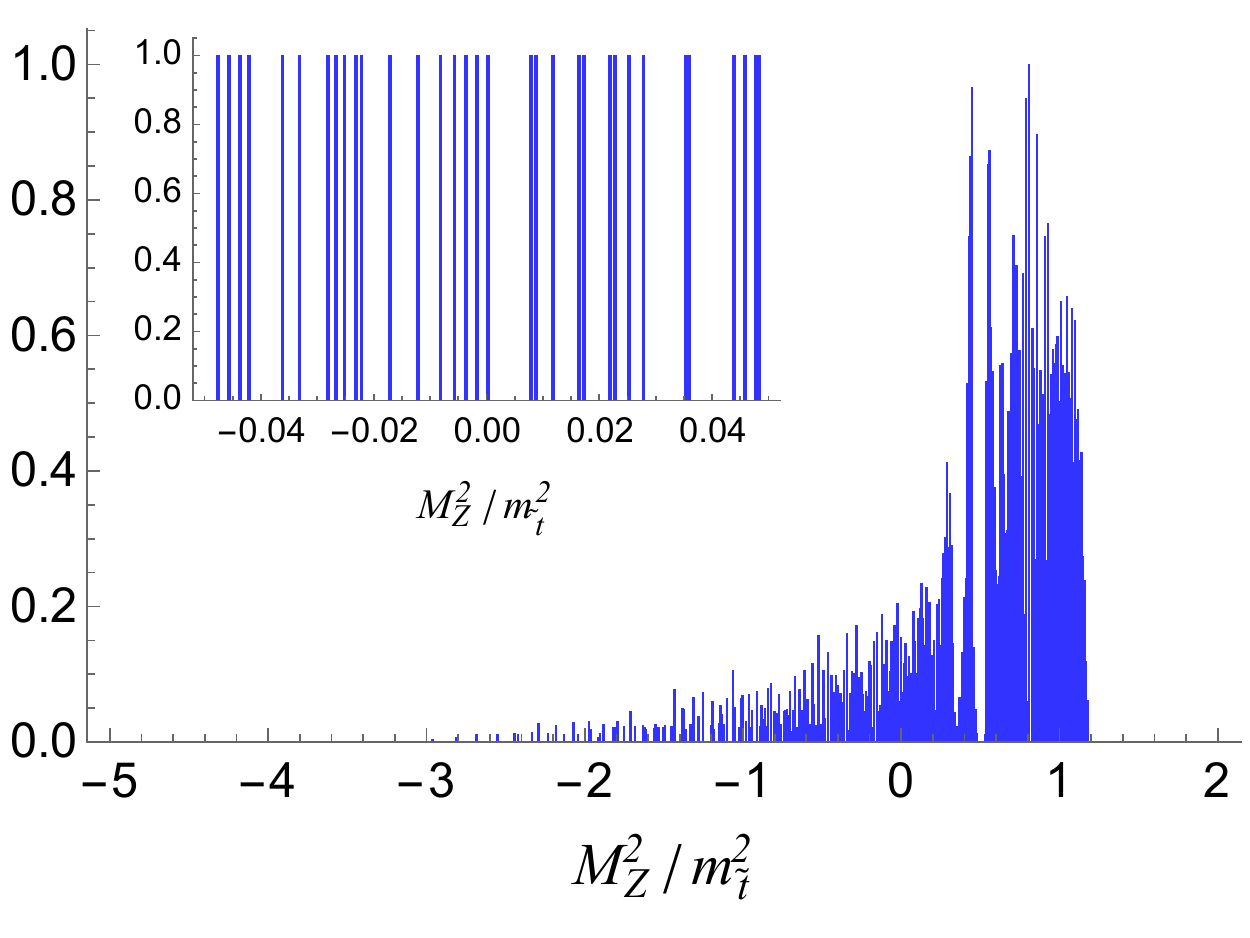}
\caption{Distribution of $M_Z^2/M_{SUSY}^2$ (left) and $M_Z^2/m_{\tilde t}^2$ (right) in the CMSSM  for  $M_{SUSY} = 3$ TeV and $\tan\beta =10$ with $A$-terms set to zero. The $\mu$, $M_{1/2}$ and $m_0^2$ are  generated in the 50\% interval around the central values set by $M_{SUSY}$  and their departures from $M_{SUSY}$ are specified by one digit.  The inset zooms in the region near zero.} 
\label{fig:CMSSM_A0}
\end{figure}

Let us start with the CMSSM with $A$-terms set to zero. The three nonzero model parameters, $\mu$, $M_{1/2}$ and $m_0^2$, are  generated in the 50\% interval around the central values set by $M_{SUSY}$  and their departures from $M_{SUSY}$ are specified by one digit. The distributions of $M_Z^2/M_{SUSY}^2$ and $M_Z^2/m_{\tilde t}^2$  for  $M_{SUSY} = 3$ TeV and $\tan\beta =10$ are shown in Fig.~\ref{fig:CMSSM_A0}. Away from the edges of the distribution, the largest gap in $M_Z^2/M_{SUSY}^2$ is about $\lesssim 0.02$ and the gap size is fairly uniform in the whole range while the largest gap size in $M_Z^2/m_{\tilde t}^2$  is about 0.008  in a large range near zero and 0.002 in a large range near the peak. The gap size in the $M_Z^2/m_{\tilde t}^2$ distribution is not uniform since it is the ratio of two distributions unlike the  $M_Z^2/M_{SUSY}^2$.
 The overall shape of the $M_Z^2/m_{\tilde t}^2$ distribution and the gap density can be understood from approximate analytic formulas that can be found for example in Ref.~\cite{Dermisek:2009si}. Solving the 1-loop RG equations and expressing the EW scale values of parameters entering Eq.~(\ref{eq:MZ}) in terms of GUT scale boundary conditions we have:
 \begin{equation}
   M_Z^2 \simeq -1.9\mu^2 + 5.9M_{1/2}^2 + 0.3m_{0}^2 
 \end{equation}
and similarly
 \begin{equation}
 m_{\tilde{t}}^2 \simeq 5.0M_{1/2}^2 + 0.6m_{0}^2.
 \end{equation}
For equal values of all GUT scale parameters we find $M_Z^2/m_{\tilde t}^2 \simeq 0.8$ which corresponds to the position of the peak in the distribution resulting from scanning model parameters around equal central values. As already mentioned, choosing different central values moves the position of the peak arbitrarily.

 Based on the gap size near the origin we conclude that, in our procedure, the CMSSM with $\mu, \; M_{1/2} ,\;m_0 \simeq M_{SUSY}$ and zero $A$-terms can generate little hierarchy $M_Z \gtrsim 0.09 m_{\tilde t}$.   An order of magnitude hierarchy can be obtained by choosing only one digit of model parameters.
 However, as already mentioned in the previous section, the position of the peak can be easily adjusted by shifting one of the parameters. For example, increasing the $\mu$-term shifts the peak and the whole distribution to smaller values and somewhat larger hierarchy can be generated.

Before we include $A$-terms, it is worthwhile to note the difference between the model we just discussed and the model with $m_0$ set to zero discussed in the previous section. The distribution in Fig.~\ref{fig:CMSSM_A0}  (left) already resembles the shaded distribution in  Fig.~\ref{fig:mu-M12} obtained with arbitrarily fine tuned inputs. Adding a new parameter  that is crudely scanned with one digit has a very similar effect to tuning the input parameters more in the model that does not contain  this parameter. A small outcome, let us say $M_Z = 0.1 m_{\tilde t}$, that requires specifying the input parameters with 2 digits in the model with two parameters  is completely ordinary in the model with one more parameter.

\begin{figure}[t]
\centering
\includegraphics[width=2.6in]{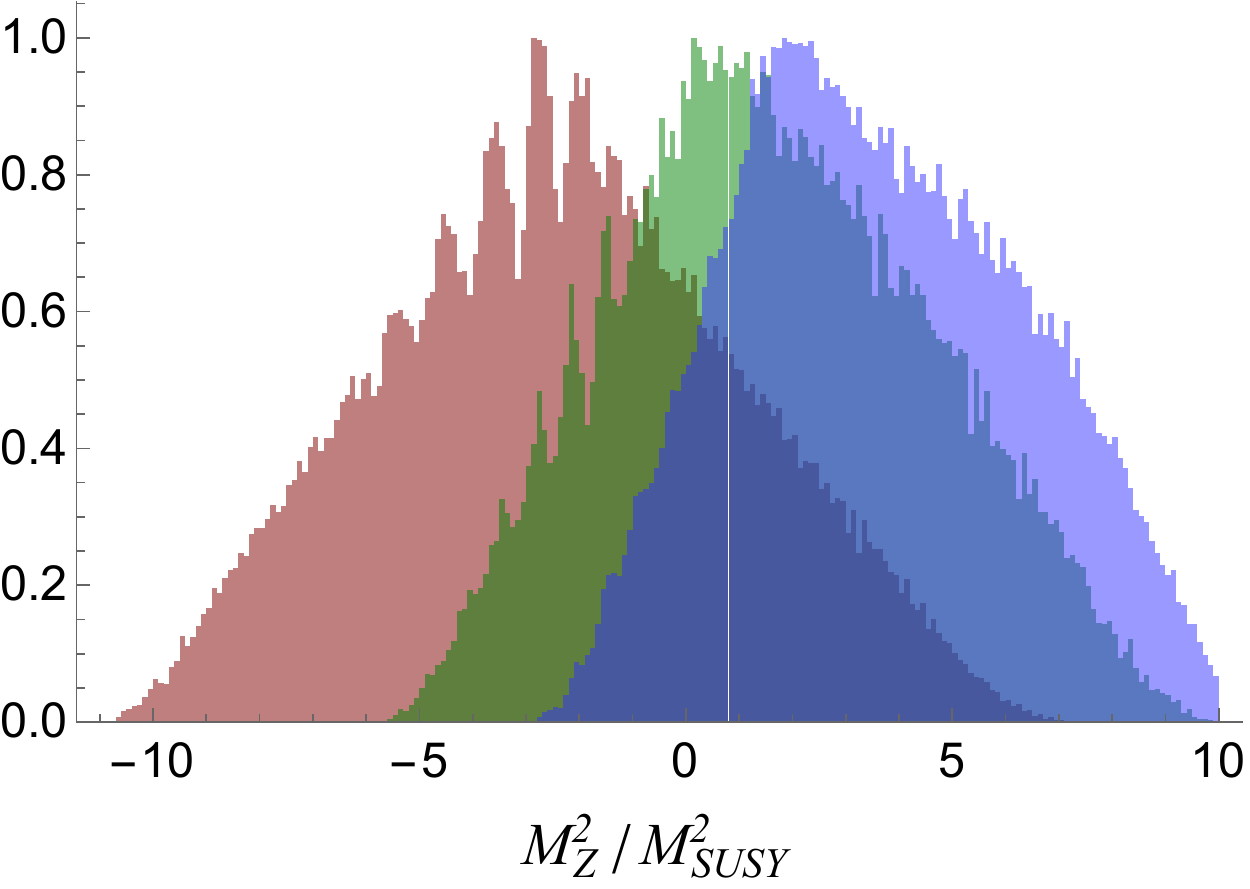}\includegraphics[width=2.6in]{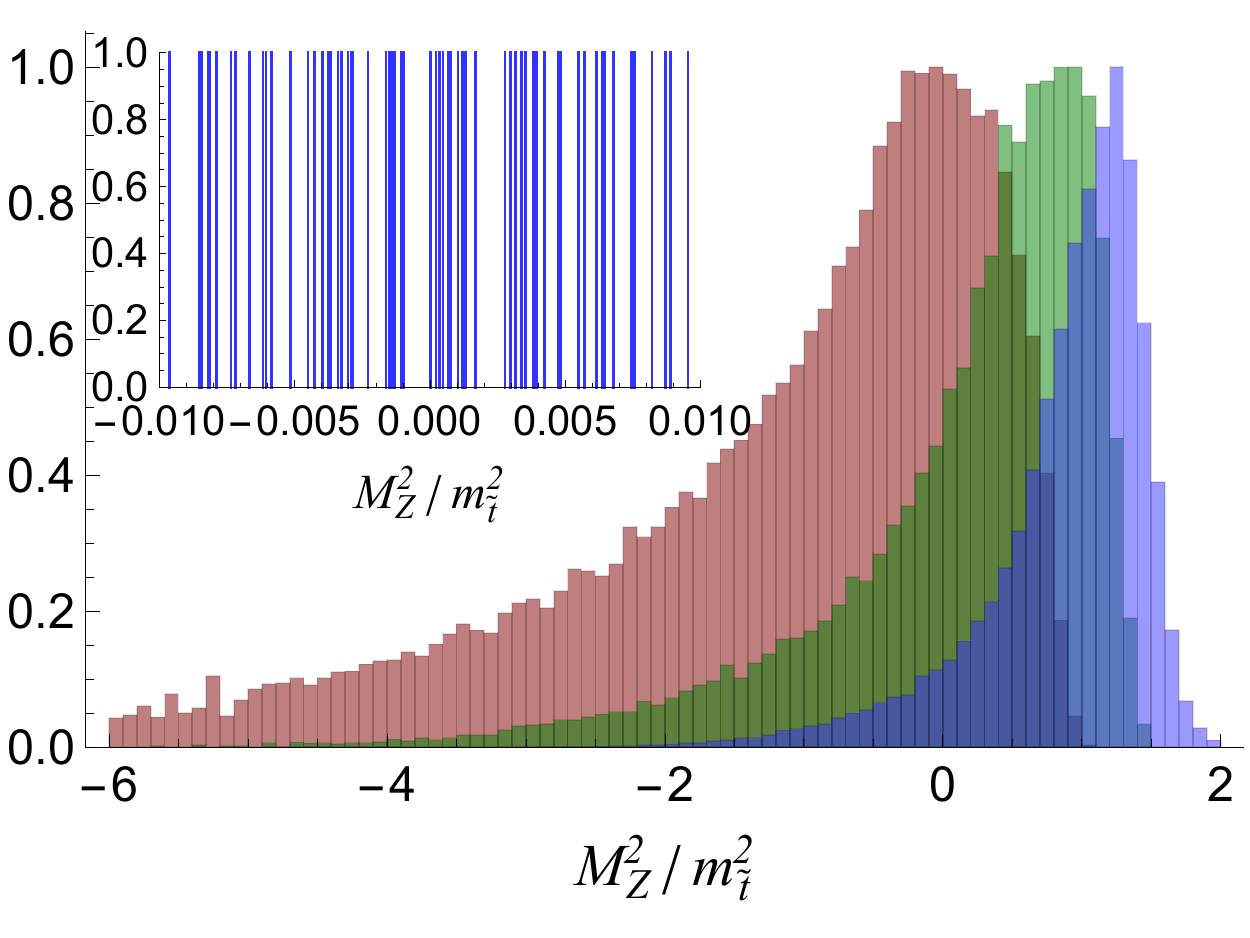}
\caption{Distribution of $M_Z^2/M_{SUSY}^2$ (left) and $M_Z^2/m_{\tilde t}^2$ (right) in the CMSSM  for  $M_{SUSY} = 3$ TeV and $\tan\beta =10$. Distributions on the right (blue) corresponds to $\mu$, $M_{1/2}$, $-A_0$ and $m_0^2$ being  generated in the 50\% interval around the central values set by $M_{SUSY}$ with their departures from $M_{SUSY}$ specified by one digit. The inset zooms in the region near zero. The distributions in the middle (green) and left (red) correspond to shifted central value of $\mu$ to $\sqrt{2}M_{SUSY}$ and $2M_{SUSY}$.} 
\label{fig:mA0}
\end{figure}

Including the universal soft trilinear couplings, $A_0$, of order  $M_{SUSY}$ with both signs, does not change the shape of the distribution significantly besides slightly spreading it, smoothing it out and filling the gaps further. Nevertheless, in order to illustrate the previous point, we show the distributions of $M_Z^2/M_{SUSY}^2$ and $M_Z^2/m_{\tilde t}^2$ now including $-A_0$ generated in the same way as $\mu$ or $M_{1/2}$ again for  $M_{SUSY} = 3$ TeV and $\tan\beta =10$ in Fig.~\ref{fig:mA0}, the distribution on the right (blue). In addition,  we also show similar distributions with shifted central values of $\mu$ to $\sqrt{2}M_{SUSY}$ (green) and $2M_{SUSY}$ (red). The distributions for the opposite sign of $A_0$ look very similar and we do not show them.

Zooming in we find that, away from the edges of distributions, the largest gap in $M_Z^2/M_{SUSY}^2$ is smaller than 0.003 (0.004) for positive (negative) sign of $A_0$ and the gap size varies very little through the whole range.  In $M_Z^2/m_{\tilde t}^2$ distribution, the largest gap size is about 0.001  in a large range near zero and near the peak for positive sign of $A_0$ while for negative sign of $A_0$ it ranges from about  0.002 in a large ranges near zero, shown in the inset of Fig.~\ref{fig:mA0}, to 0.0007 near the peak of the distribution. Increasing the size of the $\mu$ term shifts the position of the peak  and for both cases shown in Fig.~\ref{fig:mA0} the largest gap size in a large region near zero is about 0.001.

Based on the gap size near zero we conclude that the CMSSM with $\mu, \; M_{1/2} , \; \pm A_0, \;m_0 \simeq M_{SUSY}$ can result in a hierarchy $M_Z \gtrsim 0.03 m_{\tilde t}$ without the need to specify any of the model parameters by more than one digit. Larger hierarchy can be achieved with model parameters specified more precisely.

\begin{figure}[t]
\centering
\includegraphics[width=2.65in]{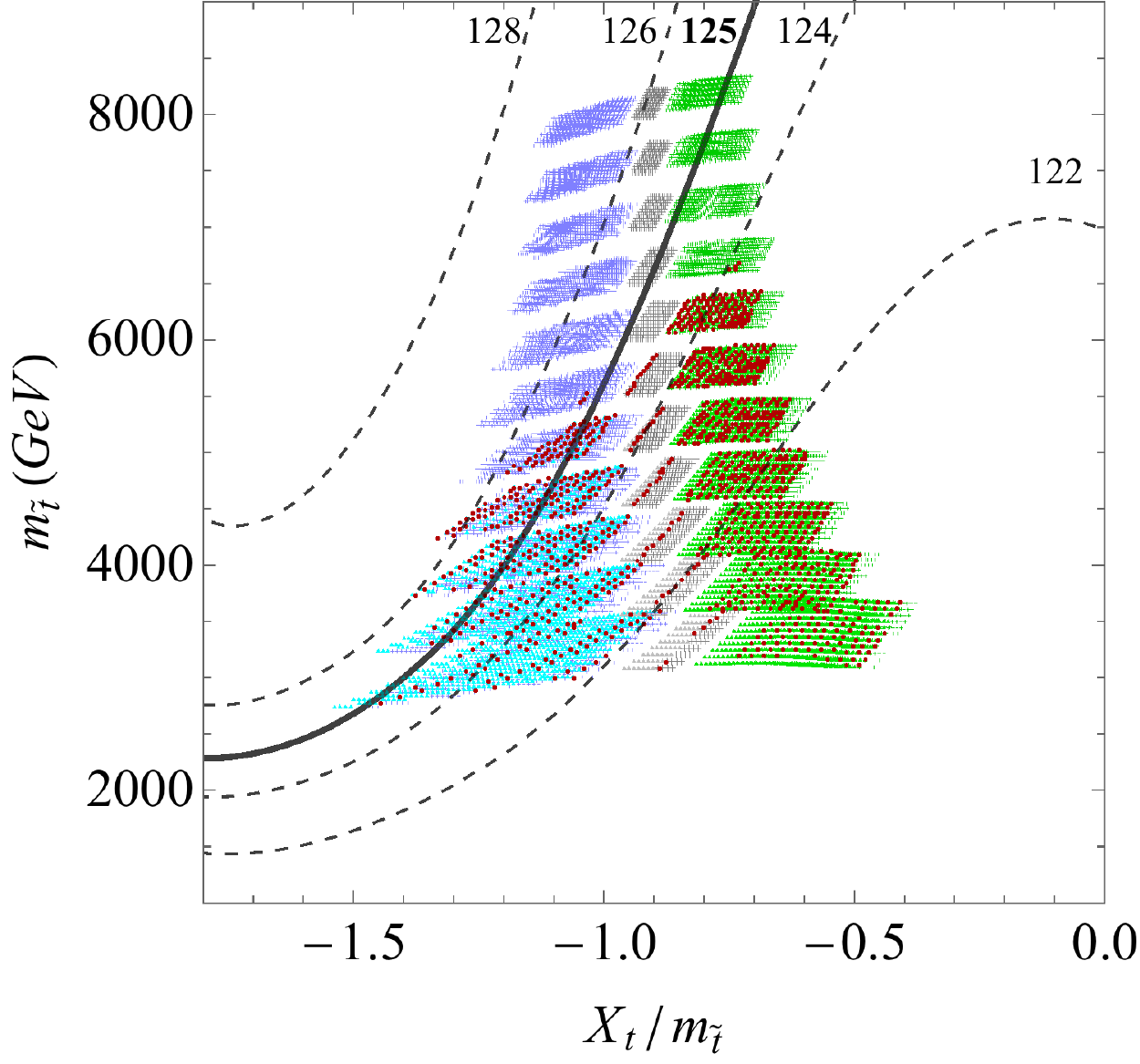}\includegraphics[width=2.5in]{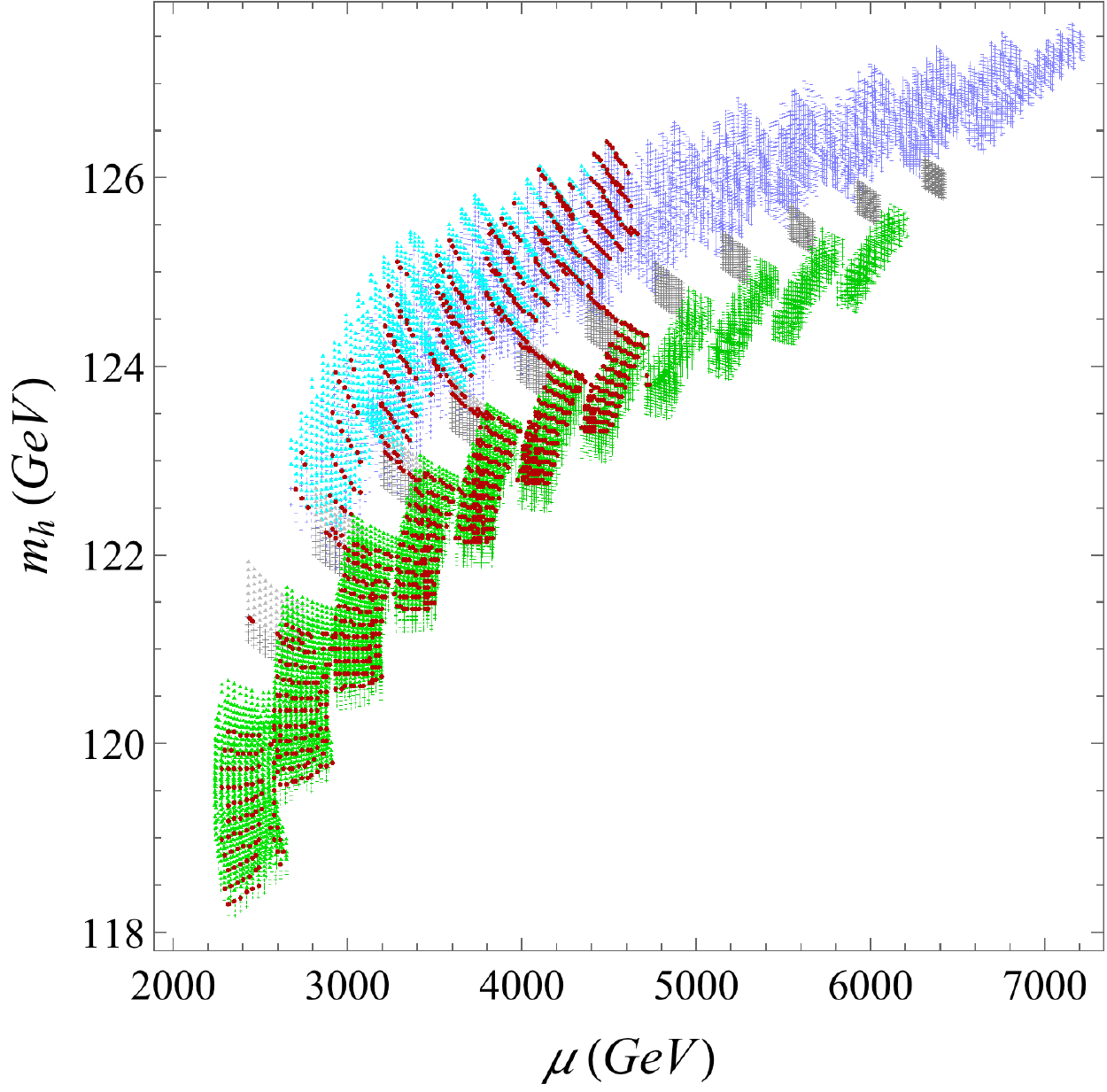}
\caption{Left: CMSSM scenarios for  $M_{SUSY} = 3$ TeV and $\tan\beta =10$ generated in our procedure with negative $A_0$ (left, blue), $A_0 = 0$ (middle, gray) and positive $A_0$ (right, green) in the $m_{\tilde t} - X_t/m_{\tilde t}$ plane. Darker shade $+$ is used for points with $M_Z^2/m_{\tilde t}^2 > 0.1$, lighter shade triangle for  $M_Z^2/m_{\tilde t}^2 < -0.1$ and the points with $|M_Z^2/m_{\tilde t}^2|< 0.1$ in each distribution are highlighted (red) dots. Contours of constant Higgs boson mass are indicated for all the remaining parameters set to 3 TeV  at $m_{\tilde t}$ (gaugino masses are set to satisfy GUT scale universality condition that would lead to $M_{\tilde g} = 3$ TeV). Right: The same scenarios with negative $A_0$ (top, blue), $A_0 = 0$ (middle, gray) and  positive $A_0$ (bottom,  green) in the $m_h - \mu$ plane with the $\mu$ term determined by the correct $M_Z$. Shades and symbols used correspond to those in the original distributions in the left plot.} 
\label{fig:higgs}
\end{figure}

Besides the hierarchy, it is also of interest to see where different CMSSM scenarios stand with respect to the Higgs boson mass. For this, the two most important parameters are the stop mass and the mixing in the stop sector given by $X_t = A_t - \mu/\tan \beta$. In Fig.~\ref{fig:higgs} (left), we plot the generated points for the three CMSSM scenarios with negative $A_0$, corresponding to the right distribution in Fig.~\ref{fig:mA0}, with $A_0 = 0$,  corresponding to the distribution in Fig.~\ref{fig:CMSSM_A0}, and with positive $A_0$ in the $m_{\tilde t} - X_t/m_{\tilde t}$ plane. As a guide, we overlay the contours of constant Higgs boson mass in $\pm 3$ GeV range from the measured value (indicating theoretical uncertainty) for all remaining parameters including the $\mu$-term set to  3 TeV at the scale of $m_{\tilde t}$ (except for gaugino masses which are assumed to satisfy the GUT scale universality condition that would lead to $M_{\tilde g} = 3$ TeV). The Higgs boson mass is calculated with  {\it FeynHiggs2.13.0-beta}~\cite{FeynHiggs}. Similar results are obtained by the effective potential method~\cite{Draper:2013oza}.

It should be noted that,
since we did not require that the correct EW scale is achieved as a result of the EWSB,  the Higgs contours are just an indication of the Higgs boson mass for a given point if the  $\mu$ parameter was adjusted to get the correct electroweak scale. We saw in Fig.~\ref{fig:mA0} that this is possible for mild changes in the $\mu$ parameter which in turn would not change the Higgs boson mass significantly. Thus the contours represent a  fair estimate. Moreover, for highlighted points (red dots) there are minimal changes in any of the parameters needed  and the Higgs boson mass is quite accurate up to theoretical uncertainties.

Alternatively, for each point we can  adjust the $\mu$-term to get the correct EWSB and calculate the Higgs boson mass with all the inputs that correspond to the given point. In Fig.~\ref{fig:higgs} (right) we plot the points in the $m_h - \mu$ plane which   shows  the size of the $\mu$ term  needed. In order to keep the  connection with previous distributions we keep the color coding as in Fig.~\ref{fig:higgs} (left).

Putting together results for the hierarchy, namely $M_Z \gtrsim 0.03 m_{\tilde t}$,  and constraint from the Higgs boson mass we see that for the  CMSSM with all the parameters of order $M_{SUSY}$, specified with one digit, only the case with negative $A_0$ can lead to sufficiently heavy Higgs boson. The stop masses needed are close to 3 TeV. Allowing more negative $A$-terms would require smaller hierarchy. On the other hand, the cases with $A_0 = 0$ or positive require larger hierarchy that can  be  obtained by specifying the model parameters more precisely.

\subsection{Non-universal Higgs or gaugino masses}

Among the simplest models studied beyond the CMSSM are the models with non-universal Higgs or non-universal gaugino masses. Both models have six parameters,  $\mu$, $M_{1/2}$, $A_0$, $m_0^2$, $m_{H_u}^2$ and $m_{H_d}^2$ in the model with non-universal Higgs masses and $\mu$, $M_{1}$, $M_{2}$, $M_{3}$, $A_0$ and $m_0^2$ in  the model with non-universal gaugino masses. These GUT scale parameters are generated in the 50\% interval around the central values set by $M_{SUSY}$  and their departures from $M_{SUSY}$ are specified by one digit. We consider both signs of $A_0$ and also the  versions of both models with $A$-terms set to zero. Since the new parameters play a sub-leading role in the EWSB, the overall shape of the distributions are very similar to the CMSSM case and thus we do not show them. Both models also perform similarly with respect to the gap size through out the distributions with differences not exceeding a factor of 2 and thus we quote the numbers  only for the model with non-universal Higgs masses. 

The largest gap in $M_Z^2/M_{SUSY}^2$ distribution near zero is  about $ 3\times 10^{-5}$ ($ 5\times 10^{-5}$) for positive (negative)  sign of $A_0$ and  $ 4\times 10^{-4}$ in the model with $A_0 = 0$. The gap size varies very little through the whole range. Similarly, the largest gap in the  $M_Z^2/m_{\tilde t}^2$ distribution in a large range near zero   is about $1 \times 10^{-5}$ for positive $A_0$,  $3 \times 10^{-5}$ for negative $A_0$ and $1.5 \times 10^{-4}$ for $A_0 = 0$. The largest gaps get smaller by about a factor of 3 near the peaks of distributions for negative $A_0$  or $A_0 = 0$ and remain almost the same for positive $A_0$.

Smaller maximal gap sizes by roughly two orders of magnitude in these models compared to corresponding versions of the CMSSM are expected since these models add two more parameters that contribute at sub-leading levels and thus  they do not spread much the distribution of outcomes. At the same time,  their contributions are sufficiently large compared to the maximal gap size in the corresponding CMSSM version so that specifying each one of them by one digit approximately splits the gaps into ten.

Based on the gap size near the origin we conclude that models with non-universal Higgs or gaugino masses with all the parameters $ \simeq M_{SUSY}$ and zero $A$-terms can generate the little hierarchy $M_Z \gtrsim 0.01 m_{\tilde t}$ with all the parameters specified by one digit. Including the $A$-terms with positive (negative) sign can increase the hierarchy up to $M_Z = 0.003 m_{\tilde t}$ ($M_Z = 0.005 m_{\tilde t}$). 

Plots related to the Higgs boson mass for these models look very similar to the CMSSM plots in Fig.~\ref{fig:higgs}. Note, that larger choices of $M_{SUSY}$ would result in shifting the points  up in Fig.~\ref{fig:higgs} (left) while only slightly changing the Higgs contours. We see that for the $A_0 = 0 $ case the stop masses required from the measured value of the Higgs boson mass are of order 6 TeV and for positive $A_0$ about 8 TeV. Such hierarchy was not possible to generate in the CMSSM with parameters specified with one digit, but in models with  non-universal Higgs masses or non-universal gaugino masses the stops can be  up to 10 TeV for $A_0 = 0 $ and 30 TeV for positive $A_0$. 
Thus, in these models, the required hierarchy between stop masses and the EW scale needed to generate the measures Higgs boson mass is possible to achieve irrespectively of the sign of the $A_0$ term, or even if $A$-terms are absent at the GUT scale, with all the model parameters specified with one digit.

\section{Discussion: ordinary vs. likely}
\label{sec:discussion}

In previous sections, we have identified the smallest outcome in a given model that is guaranteed to occur in the  distribution with model parameters selected in 10\% steps around comparable central  values. In Sec.~\ref{sec:method} we argued that such an outcome is a completely ordinary outcome that cannot be disfavored by any logical argument compared to any other outcome resulting from selecting model parameters in given steps.
Since this is somewhat counterintuitive and it seemingly contradicts common probabilistic arguments related to fine tuning, let us discuss it further.

The probabilistic argument related to a small outcome being unnatural is related to the fact that there are more large numbers compared to small numbers.  For example, 
let us suppose that one integer between 1 and 1000 is randomly selected.\footnote{Normalizing the largest entry to 1, this is the same as selecting one number from: 0.001, 0.002, ..., 1.00.} The analogy with common naturalness arguments for EW symmetry breaking is saying that an outcome smaller than 100 has only a 10\% probability, an outcome smaller than 10 has a probability of just 1\% and the number 1 has a probability  0.1\% of being selected. Therefore, if one adopts a 10\%  ``naturalness" criterion,  the outcome should  be somewhere between 100 and 1000 (or 90\% of randomly selected outcomes will be in this range).  Although this probabilistic statement is correct, it actually has no value in predicting the outcome. Clearly, nobody should be surprised if any number between 1 and 1000 is picked, even if it is number 1. The small probability of number 1 being picked does not matter since it is the same as the probability of any other outcome.\footnote{Comparing the probability of a small specific outcome with the probability of an outcome in some large range, as is commonly done, is misleading.  Moreover such an argument can be reversed and could be used to claim that the largest numbers are not natural, e.g. anything larger than 900 has only 10\% chance. Or it could be used to disfavor outcomes right in the middle or anywhere else.}  If only one number is selected, there is no logical argument one can make to disfavor number 1 compared to any other number. 

The seemingly contradicting connection between the probability of a given outcome and the level of tuning of model parameters required is there only for models with two parameters.\footnote{In a model with only two parameters contributing to a given observable the probabilistic argument indeed gives the correct level of fine tuning required. For example, in order for two ${\cal O}(1)$ contributions to a given observable to produce an outcome of order 0.001, we must tune one of the parameters at 0.1\% level. In a random probabilistic scan such an outcome would appear with 0.1\% probability. If we insisted on not tuning model parameters by more than 10\% the smallest expected outcome would be of order 0.1 and would appear with 10\% probability in a random scan. However, there is no contradiction with the previous paragraph. In our approach, the smallest outcome  guaranteed to appear in any distribution from two parameters selected in 10\% steps  would  also be of order 0.1.} As we saw, with more parameters contributing to a given observable, a small outcome does not necessarily require careful choices of model parameters. For example, in the CMSSM with 4 parameters selected in 10\% steps an outcome for $M_Z^2/m_{\tilde t}^2$ as small as 0.001 will appear in any distribution no matter how model parameters are shifted from central values. Once such an outcome is guaranteed to appear in the distribution, there is no logical argument one can make, just like for the random integer example above,  to disfavor it compared to any other outcome. A completely ordinary outcome does not have to be likely. With more parameters contributing, the probability of any specific outcome is decreasing. The small probability is not a sign of fine tuning required but rather an indication  that a model with more parameters is less predictive.


\section{Conclusions}
\label{sec:conclusions}

	In this paper, we have adopted an approach to study the possible hierarchy between the electroweak scale and the scale of superpartners, $M_Z^2/M_{SUSY}^2$ or $M_Z^2/m_{\tilde t}^2$, that can be obtained without specifying model parameters by more than one digit (or with better than 10\% precision).  The smallest such outcome is estimated from the largest gap  in the distribution obtained from selecting model parameters  in 10\% steps around a common scale. Since an outcome at least as small as the largest gap size is guaranteed to occur in any distribution with the model parameters selected in 10\% steps around randomly shifted central values, with the same probability as any other individual outcome, it is completely ordinary.

We have found  that the maximal hierarchy between the electroweak scale and stop masses, given  that model parameters are not specified beyond one digit, ranges from a factor of $\sim 10-30$ for  the CMSSM (depending on the sign and size of $A_0$) up to  $\sim 300$ for models with non-universal Higgs or gaugino masses when all non-zero parameters are of the same order, $M_{SUSY}$, at the GUT scale. For the  CMSSM scenarios only those with negative $A$-terms, $-A_0 \simeq M_{SUSY}$, can lead to sufficiently heavy Higgs boson with the little hierarchy generated by minimally specified model parameters. In this case, it is possible to obtain $\sim 100$ GeV  electroweak scale  and $\sim 3$ TeV stop masses needed for the Higgs mass by choosing only one digit of model parameters no matter what the remaining digits are. In models with non-universal Higgs or gaugino masses, the required hierarchy between stop masses and the EW scale needed to generate the measures Higgs boson mass is possible to achieve with minimally specified parameters irrespectively of the sign of the $A_0$ term, or even if $A$-terms are absent at the GUT scale.

 In Sec.~\ref{sec:discussion}, we argued that the probabilistic arguments typically employed to estimate the hierarchy indicate the required level of tuning only for models with two parameters.  The 10\% interval of the smallest outcomes based on probabilistic arguments (or 10\% tuning requirement based on sensitivity measures)  is equivalent to specifying model parameters with 10\% precision in models with just two parameters.   However, we saw that in models with more than two parameters, a smaller outcome that would normally be disfavored by probabilistic arguments or sensitivity measures can be obtained with the same level of precision in model parameters and the possible little hierarchy grows with the complexity of the model. For example, an outcome for $M_Z^2/m_{\tilde t}^2$ as small as 0.001 in the CMSSM with 4 parameters would appear in any distribution where model parameters are selected in 10\% steps around a common scale. Since an outcome of that size is guaranteed to appear, regardless of ${\cal O}(1)$ shifts in the parameters, it is completely ordinary and cannot be disfavored compared to any other individual outcome.   A small probability of the smallest completely ordinary outcome is not problematic since any other outcome obtained from given selection method has the same probability.

The little hierarchy suggested by the Higgs boson mass (besides direct experimental limits) might not be a sign of a large fine tuning required among the model parameters  but rather an indication of  complexity of the correct model of nature.


\vspace{0.5cm}
\noindent
{\bf Acknowledgments:} This work was supported in part by the U.S. Department of Energy under grant number {DE}-SC0010120.



\end{document}